\documentclass[a4paper,11pt]{article}
\pdfoutput=1 

\usepackage{jcappub} 

\usepackage[T1]{fontenc} 

\title{\boldmath Gravitational waves from the E-model inflation with Gauss-Bonnet correction}

\author[a,1]{Tie-Jun Gao,\note{Corresponding author.}}
\author[b]{Jian-Xia Guo,}

\affiliation[a]{School of Physics, Xidian University,\\Xi'an 710071, China}
\affiliation[b]{Shangluo University,\\Shangluo 726000, China}

\emailAdd{tjgao@xidian.edu.cn}
\emailAdd{guojianxia1986@qq.com}

\abstract{In this work, we study the generation of gravitational waves in the E-model inflation with the scalar field non-minimally coupled to the Gauss-Bonnet term. Considering a wall-crossing behavior in the moduli space, we parameterize the coupling coefficient $\xi$ as a step-like function, then if $V_{,\phi}\xi_{,\phi}>0$, the Gauss-Bonnet term dominate the inflation  dynamics, causing a short rapid-decline phase of the inflaton, and for appropriate parameter spaces, the mode equation of tensor perturbations develops a transient growing solution. This process generates a peak in the tensor perturbation power spectrum, corresponding to a peak in the gravitational wave energy spectrum around the nanohertz  frequency band.
Further more, we investigate the feasibility of generating double peaks in the gravitational wave spectrum using a double-step coupling, For certain parameter choices, one peak lies near nanohertz frequencies, while the other is around millihertz  frequencies. Consequently, these gravitational waves  can be observed by the pulsar timing array and the space-based gravitational wave detectors such as LISA, simultaneously.}

\begin{document}
\maketitle
\flushbottom

\section{Introduction}
\label{sec:intro}
In 2015, the LIGO and Virgo collaborations observed a gravitational wave (GW) signal generated by the merger of two black holes\cite{LIGOScientific:2016aoc}, marking the beginning of GW astronomy.
Besides these from mergers, another source of GWs is the perturbation  of inflation in the early universe. Recently, in 2023, the worldwide pulsar timing array (PTA) collaborations, such as the North American Nanohertz Observatory for Gravitational Waves(NANOGrav)\cite{NANOGrav:2023gor,NANOGrav:2023hde,NANOGrav:2023hvm}, the European Pulsar Timing Array (EPTA) \cite{EPTA1,EPTA2}, the Chinese Pulsar Timing Array (CPTA)\cite{CPTA1} and the Parkes Pulsar Timing Array (PPTA) \cite{PPTA1,PPTA2}  announced that a stochastic GW backgrounds has been observed with the frequency around nanohertz, which may come from the  early universe.

However, the CMB scales observations give a strong constraint on the tensor-to-scalar ratio $r<0.064$  by the Planck 2018 data in combination with BICEP2/Keck Array\cite{Planck:2018jri}, which is too small to be observed in the near future. Whereas the observations of the CMB only give constraints at high energy scales, if the power spectrum of tensor perturbations is enhanced at low energy scales, the GWs can be detected~\cite{ref10,ref11,ref12,ref13,ref14,ref15,ref16,ref17,ref171,ref172,ref173,ref174,ref175}.

On the other hand, although the inflation of the early universe has been established by a large number of observations, the origin of inflation is not entirely clear. An interesting approach is to study inflationary models within the framework of quantum gravity theories such as superstring theory, and a low-energy effective  theory is to introduce a non-minimum coupling between the inflaton $\phi$ and the Gauss-Bonnet(GB) term in the action\cite{GB1,GB2,GB3},see also Ref.\cite{GB4,GB5,GB6,Guo2009,Guo:2010jr,Guo2013}. So in this work, we discuss the effects of the GB coupling in the framework of E-model attractor inflation. Assuming a wall-crossing process of $\phi$ in the moduli space,  the coupling coefficient $\xi(\phi)$ will have the form of a step-like function\cite{step1,step2,Kawai:2023nqs,Kawai:2021edk}. If $V_{,\phi}\xi_{,\phi}<0$, this  leads to a de Sitter fixed point and thus the inflaton undergos an ultra-slow-roll phase. The related scalar-induced GWs have been discussed in \cite{Kawai:2021bye,Kawai:2021edk,Zhang:2021rqs,Solbi:2024zhl}. In contrast, if $V_{,\phi}\xi_{,\phi}>0$,   when the inflaton rolls close to the step point,  it  undergos a brief acceleration phase. Notably, if $c^2_T$  becomes negative for a short time during this accelerating phase, the mode equation for tensor perturbations develops a transient growing solution, this results in a significant peak in the tensor power spectrum, corresponding to a peak in the present-day GW spectrum.

In addition, if the moduli space is assumed to have multiple minima, then there can be two or more wall-crossing processes of $\phi$. So we assume that the coupling coefficient $\xi(\phi)$ has the form of a double-step function, which will result in  a double-peak GW spectrum. For some parameter sets, the GWs can be observed at the nanohertz frequency band by the PTA  observation and at the millihertz frequency band by space-based GW detectors such as LISA and Taiji, simultaneously.

The paper is organized as follows. In the next section, we set up the inflationary model with GB correction. In Sec.3, we list the relevant formulas of scalar and tensor perturbations, and then give the numerical results of the inflation dynamics for the model with a single-step function coupling. The calculations of the corresponding GW energy spectrum are presented in Sec.4. In Sec.5, we discuss the possibility of generating a GW spectrum with double peaks using the double-step function coupling, and the numerical results are also shown there.  The last section is devoted to summary.

\section{The model}


We consider the following action\cite{Guo:2010jr}
\begin{eqnarray}
&&S = \int d^4x\sqrt{-g}
\left[\frac12 R - \frac{1}{2}(\nabla \phi)^2
 - V(\phi)-\frac12\xi(\phi) R_{\rm GB}^2 \right],
\label{action}
\end{eqnarray}
with the scalar field $\phi$ non-minimum coupling to the GB term $R_{\rm GB}^2=R_{\mu\nu\rho\sigma} R^{\mu\nu\rho\sigma} - 4 R_{\mu\nu} R^{\mu\nu} + R^2$, and $\xi(\phi)$ is the coupling coefficient. We work in Planckian units, $\hbar=c=8 \pi G=1$.
The scalar potential is assumed to have the form of E-model attractor inflation\cite{Kallosh:2013daa,Kallosh:2013hoa,Kallosh:2013yoa}
\begin{eqnarray}
&&V(\phi) = V_0 \Big[1-\exp(-\sqrt{\frac{2}{3\alpha}}\phi)\Big]^{2n},
\end{eqnarray}
with $V_0$, $\alpha$ and $n$ are three parameters.

After varying the action in the Friedmann-Robertson-Walker(FRW) homogeneous background, one can obtain the Friedmann equation and the scalar field
equation as below
\begin{eqnarray}
\label{beq1}
&& 6H^2 =  \dot{\phi}^2 + 2V + 24\dot{\xi}H^3, \\
&&  \ddot{\phi} + 3 H \dot{\phi} + V_{,\phi} +12
\xi_{,\phi}
 H^2 \left(\dot{H}+H^2\right) = 0,
\label{beq2}
\end{eqnarray}
with a dot denotes derivative with respect to the cosmic time $t$ and $(...)_{,\phi}$ denotes derivative
with respect to the scaler field $\phi$, thus one has $\dot{\xi}=\xi_{,\phi}\dot{\phi}$. 
We can see that the contribution from the GB term comes entirely from the first derivative of $\xi(\phi)$ with respect to the scalar field $\phi$.

In the standard single-field inflation theory, it is useful to define a series of slow-roll parameters using the Hubble parameter as below~\cite{HubbleSR1,HubbleSR2,HubbleSR3}
\begin{eqnarray}
\epsilon_1 = -\frac{\dot{H}}{H^2},
 \quad \epsilon_{i+1} = \frac{d\ln|\epsilon_i|}{H d t},
 \quad i \ge 1.
\end{eqnarray}
However, due to the presence of GB coupling, the new degrees of freedom suggest to introduce additional GB slow-roll parameters\cite{Guo:2010jr}
\begin{eqnarray}
\delta_1 = 4 \dot{\xi} H,
 \quad \delta_{i+1} = \frac{d\ln|\delta_i|}{H d t},
 \quad i \ge 1.
\end{eqnarray}

The coupling coefficient $\xi(\phi)$ is chosen as a hyperbolic tangent function
\begin{eqnarray}
\xi=\xi_0\tanh[\xi_1(\phi-\phi_c)],
\end{eqnarray}
which is similar to the step function. Where $\xi_0$ and $\xi_1$ are real constants, and $\phi_c$ marks the position of the step.
Such step-like coupling coefficient comes from a wall-crossing process in the moduli space\cite{Kawai:2021edk,Kawai:2023nqs}. When the modulus $\phi$ crosses a finite thickness domain-wall at $\phi_c$ between different Bogomol'nyi-Prasad-Sommerfield (BPS) spectra\cite{Antoniadis:1992sa,Harvey:1995fq}, the behavior of $\xi$ can be modeled by the step like function(2.7).

 Since the contribution of the GB term to the background equations only comes from $\xi_{,\phi}$, therefore, in this model, when $\phi$ is far away from $\phi_c$, the contribution of the GB term can be ignored, thus the prediction at the CMB scale will be similar to the case without the GB term. However, once $\phi$ rolls near the critical point $\phi_c$, due to the large derivative of the step function $\xi_{\phi}$, the contribution from the GB term becomes particularly important.
 If $V_{,\phi}\xi_{,\phi}<0$, which will result in a nontrivial fixed point, causing the inflation to go through an ultra-slow-roll phase, and the associated generation of scalar induced GWs has been discussed in\cite{Kawai:2021edk,Kawai:2021bye,Zhang:2021rqs,Solbi:2024zhl}. In this work, we focus on the opposite case $V_{,\phi}\xi_{,\phi}>0$, and in the vicinity of $\phi_c$, the scalar field equation can be approximated as
\begin{align}
    \dot{\phi}^2 \simeq
    -H^2(1-\epsilon_1)\delta_1\,.
\end{align}
We can see that since the slow-roll parameter $\delta_1$ grows significantly as the inflaton rolls through the critical point, the velocity of $\phi$ increases significantly, that is, $\phi$ experiences a rapid decline. After that, the contribution of GB term is negligible again and the inflation returns to the usual slow-roll case. In addition, if the coefficient $c^2_T$ in the mode equation of the tensor perturbations is less than zero for a short time during the rapid decline, there will have a transient growing solution. Which will causes a peak in the tensor power spectrum, and resulting in an observable GW spectrum. We take the parameter set I of Table I as an example, and the evolution of the inflaton $\phi$  with the e-folding number $N_e$ are show in Fig.1.

\begin{table}
\centering
\begin{tabular}{|c||c|c|c|c|c|c|}
\hline
Sets&$n$&$V_0$&$\sqrt{\frac{2}{3\alpha}}$&$\xi_0$&$\xi_1$&$\phi_c$\\
\hline
I&$1/4$&$1.445\times10^{-9}$&$0.015$&$3.2\times10^{9}$&$1.0$&$7.5$\\

II&$1/4$&$2.425\times10^{-9}$&$0.010$&$0.7\times10^{9}$&$2.3$&$7.4$\\
III&$1$&$1.580 \times10^{-9}$&$0.170$&$4.6\times10^{7}$&$15.0$&$8.5$\\
IV&$1$&$1.233\times10^{-9}$&$0.200$&$4.0\times10^{7}$&$20.0$&$9.0$\\
\hline

\end{tabular}
\caption{Examples of the parameter set.}
\end{table}

\begin{figure}
\centering
 \begin{minipage}[t]{0.7\linewidth}
	\centering
	\includegraphics[width=.99\textwidth]{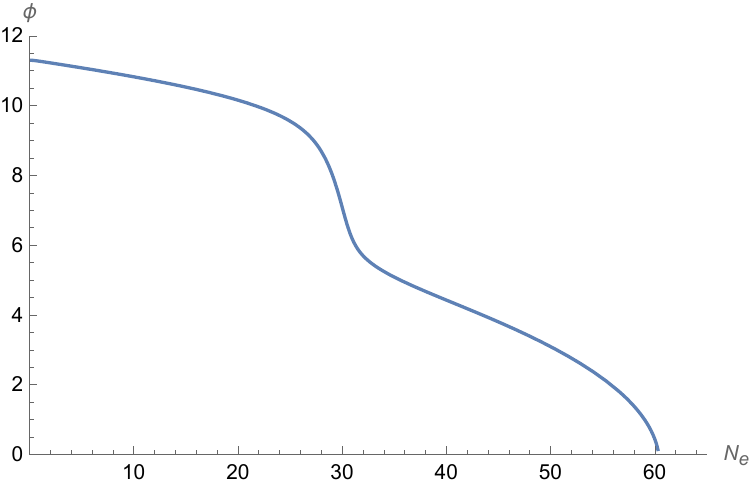}
	\label{fig:a}
 \end{minipage}
 \caption{The relation of inflaton $\phi$  evolution with the e-folding number for the parameter set I.}
\label{fig:1}
\end{figure}

\section{Perturbation power spectrum \label{sect3}}
In order to estimate the perturbation power spectrum precisely, we need to solve the Mukhanov-Sasaki(MS) equation of Fourier modes numerically.
For the scalar perturbations, the mode equation takes the following form~\cite{MS1,MS2,MS3}
\begin{eqnarray}
v_k'' + \left(c_{\mathcal{R}}^2 k^2 - \frac{z''_{\mathcal{R}}}{z_{\mathcal{R}}}\right)v_k = 0,
\label{spm}
\end{eqnarray}
where a prime represents a derivative versus the conformal time $\tau = \int a^{-1}dt$, and $v_k\equiv z_{\mathcal{R}} {\mathcal{R}}_k$ with ${\mathcal{R}}_k$ is  the Fourier mode of the  gauge invariant curvature perturbation ${\mathcal{R}}$. $z_{\mathcal{R}}$ and
the sound speed $c_{\mathcal{R}}$ can be expressed in terms of the the slow-roll parameters as\cite{Guo:2010jr}
\begin{eqnarray}
&& z_{\mathcal{R}}^2 = \frac{a^2(\dot{\phi}^2 +
 6 \Delta \dot{\xi}H^3)}{(1 - \frac12 \Delta)^2 H^2}=a^2 \frac{F}{(1-\frac12 \Delta)^2}, \\
&& c_{\mathcal{R}}^2 = 1 + \frac{8 \Delta \dot{\xi} H \dot{H} +
 2 \Delta^2 H^2 (\ddot{\xi}-\dot{\xi}H)}{\dot{\phi}^2+6\Delta\dot{\xi}H^3}=1 - \Delta^2 \frac{2\epsilon_1+\frac12 \delta_1(1-5\epsilon_1-\delta_2)}{F},
\label{cr2}
\end{eqnarray}
with
\begin{eqnarray}
&&F = 2\epsilon_1-\delta_1(1+\epsilon_1-\delta_2)+\frac32 \Delta \delta_1,\\
&&\Delta = \frac{4\dot{\xi}H}{1-4\dot{\xi}H}=\frac{\delta_1}{(1-\delta_1)}.
\label{spm}
\end{eqnarray}
Then the effective mass term in (3.1) reads
\begin{equation}\begin{aligned}
\frac{z_{\mathcal{R}}^{\prime \prime}}{z_{\mathcal{R}}}=& a^{2} H^{2}\left[2-\epsilon_{1}+\frac{3}{2} \frac{\dot{F}}{H F}+\frac{3}{2} \frac{\dot{\Delta}}{H\left(1-\frac{1}{2} \Delta\right)}+\frac{1}{2} \frac{\ddot{F}}{H^{2} F}+\frac{1}{2} \frac{\ddot{\Delta}}{H^{2}\left(1-\frac{1}{2} \Delta\right)}\right.\\
&\left.-\frac{1}{4} \frac{\dot{F}^{2}}{H^{2} F^{2}} +\frac{1}{2} \frac{\dot{\Delta}^{2}}{H^{2}\left(1-\frac{1}{2} \Delta\right)^{2}}+\frac{1}{2} \frac{\dot{\Delta}}{H\left(1-\frac{1}{2} \Delta\right)} \frac{\dot{F}}{H F}\right],
\end{aligned}\end{equation}
with
\begin{equation}\begin{aligned}
\frac{\dot{F}}{H}=& \epsilon_{1} \epsilon_{2}\left(2-\delta_{1}\right)-\delta_{1} \delta_{2}\left(1+\epsilon_{1}-\delta_{2}-\delta_{3}\right)+\frac{3}{2} \Delta \delta_{2}\left(\Delta+\delta_{1}\right), \\
\frac{\dot{\Delta}}{H}=& \Delta^{2} \frac{\delta_{2}}{\delta_{1}}, \\
\ddot{H} =&\epsilon_{1} \epsilon_{2}\left(-\epsilon_{1}+\epsilon_{2}+\epsilon_{3}\right)\left(2-\delta_{1}\right)+\epsilon_{1} \delta_{1} \delta_{2}\left(1+\epsilon_{1}-2 \epsilon_{2}-\delta_{2}-\delta_{3}\right)\\
&-\delta_{1} \delta_{2}^{2}\left(1+\epsilon_{1}-\delta_{2}-\delta_{3}\right) -\delta_{1} \delta_{2} \delta_{3}\left(1+\epsilon_{1}-2 \delta_{2}-\delta_{3}-\delta_{4}\right) \\
&+\frac{3}{2} \Delta \delta_{2}\left(\Delta+\delta_{1}\right)\left(-\epsilon_{1}+\Delta \frac{\delta_{2}}{\delta_{1}}+\delta_{3}\right) +\frac{3}{2} \Delta \delta_{2}\left(\Delta^{2} \frac{\delta_{2}}{\delta_{1}}+\delta_{1} \delta_{2}\right), \\
\ddot{\Delta} =&\Delta^{2} \frac{\delta_{2}}{\delta_{1}}\left(-\epsilon_{1}+2 \Delta \frac{\delta_{2}}{\delta_{1}}-\delta_{2}+\delta_{3}\right).
\end{aligned}\end{equation}

Similarly, the mode equation of tensor perturbations satisfy~\cite{MS1,MS2,MS3}
\begin{eqnarray}
u_k'' + \left(c_{T}^2 k^2 - \frac{z''_{T}}{z_{T}}\right)u_k = 0,
\label{tpm}
\end{eqnarray}
where $u_k$ is the Fourier mode of $u^{\pm}$, which is defined via
\begin{eqnarray}
h_{ij}=\frac{\sqrt{2}}{z_T}\sum_{\pm} u^{\pm} e_{ij}^{\pm},
\label{tpm}
\end{eqnarray}
with $e_{ij}^{\pm}$ is the polarization tensor, and ${\pm}$ indicates the two polarization models.
$z_{T}^2$ and $c_{T}^2$ can be written in terms of the Hubble and GB slow-roll parameters as
\begin{eqnarray}
&& z_{T}^2 = a^2(1 - 4\dot{\xi}H) = a^2 (1-\delta_1) , \\
&& c_{T}^2 = 1 - \frac{4(\ddot{\xi}-\dot{\xi}H)}{1 - 4\dot{\xi}H} = 1 + \Delta (1-\epsilon_1-\delta_2).
\label{ct2}
\end{eqnarray}
And the effective mass term $z''_{T}/z_{T}$ in the tensor mode equation~(3.8) is
\begin{eqnarray}
\label{ztpp}
\frac{z_{T}''}{z_{T}} &=& a^2H^2
 \bigg[2-\epsilon_1-\frac32 \Delta\delta_2-\frac12\Delta\delta_2(-\epsilon_1+\delta_2+\delta_3)-\frac14\Delta^2\delta_2^2 \bigg].
\end{eqnarray}

The evolutions of the relevant parameters $\epsilon_1, \delta_1, \delta_2$, $\Delta$ in (3.11) and the square of the tensor sound speed $c^2_T$ with the e-folding number $N_e$ are shown in Fig.2. We can see that near the step point, the parameter $\delta_1$ has a valley slightly lower than $-1$, which causes the curve of $\Delta$ to have a valley. Meanwhile, $\epsilon_1$ and $\delta_2$ have a change from positive to negative near this valley and equal zero at the bottom of the valley. These behaviors correspond to a narrow and deep valley in the curve of $c_T^2$.
\begin{figure}
 \begin{minipage}[t]{0.49\linewidth}
	\centering
	\includegraphics[width=.99\textwidth]{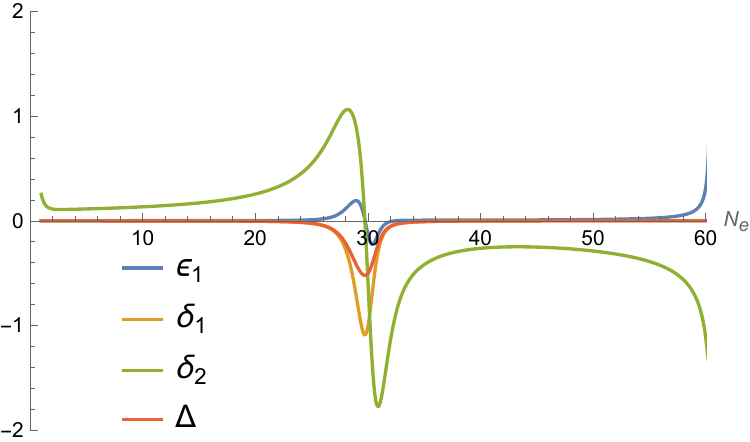}\label{fig:a} %
 \end{minipage}
 \begin{minipage}[t]{0.49\linewidth}
	\centering
	\includegraphics[width=.99\textwidth]{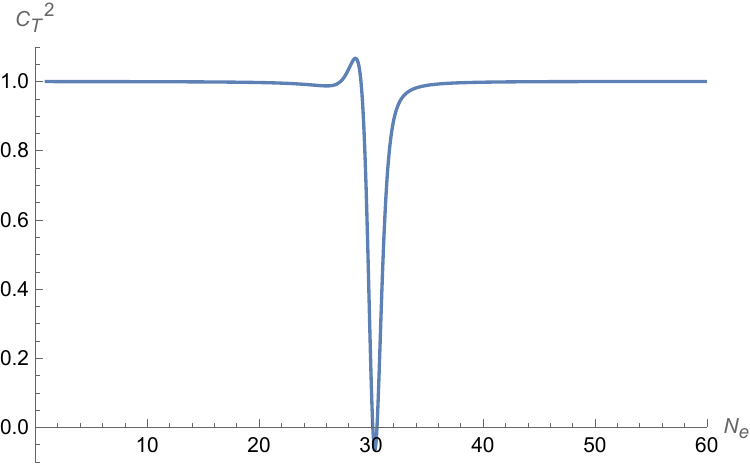}
	\label{fig:b}
 \end{minipage}
 \caption{The relation of the parameters $\epsilon_1, \delta_1, \delta_2$ and $\Delta$ (left) and $c^2_T$ (right) evolution with the e-folding number for the parameter set I.}
\label{fig:1}
\end{figure}


In addition, although the brief negativity of $c_T^2$ can cause the gradient instability of high-momentum modes, however, the model is a low-order effective theory, and if the Lagrangian contains some high-order terms, the corresponding $c_T^2$ will depend on $k$. For example, as shown in Ref.\cite{Cai:2016thi}, if the Lagrangian contains high-order terms $\propto \nabla_i R^{(3)} \nabla^i R^{(3)} $, with $R^{(3)}$ is the induced 3-dimensional Ricci scalar, which will contribute $k^4$ term in (3.11), this ensures that for sub-horizon modes with large $k$, the $k^4$ terms dominate, making the $c_T^2$ of the mode remain positive, and thus no instability occurs. Meanwhile, for small $k$, the correction of higher-order terms can be ignored, and thus the background and perturbation  equations will not be affected.

Considering the initial conditions of the Bunch-Davies type~\cite{BD},
\begin{align}
\lim _{\tau \rightarrow-\infty} v_k(\tau) & =\frac{1}{\sqrt{2 c_{\mathcal{R}} k}} e^{-i c_{\mathcal{R}} k \tau}, \\
\lim _{\tau \rightarrow-\infty} u_k(\tau) & =\frac{1}{\sqrt{2 c_T k}} e^{-i c_T k \tau},
\end{align}
the power spectrum of the scalar and tensor perturbations can be calculated by
\begin{align}
\mathcal{P}_R & =\frac{k^3}{2 \pi^2} \left|\frac{v_k}{z_{\mathcal{R}}}\right|^2, \\
\mathcal{P}_T & =\frac{k^3}{\pi^2} \left|\frac{2 u_k}{z_{T}}\right|^2.
\end{align}

The numerical results of the tensor perturbation power spectrum  for the four parameter sets are shown in Fig.3.
\begin{figure}[htb]

 \begin{minipage}[t]{0.49\linewidth}
\centering
  \includegraphics[width=.99\textwidth]{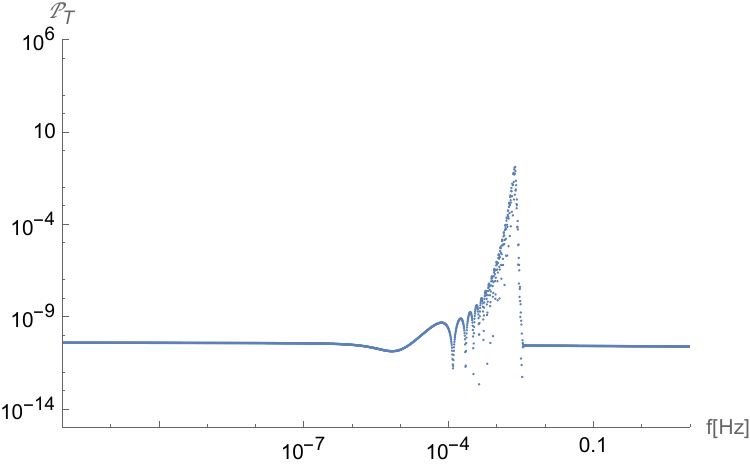}\hfill
  \centerline{I}
 \end{minipage}
 \begin{minipage}[t]{0.49\linewidth}
	\centering
  \includegraphics[width=.99\textwidth]{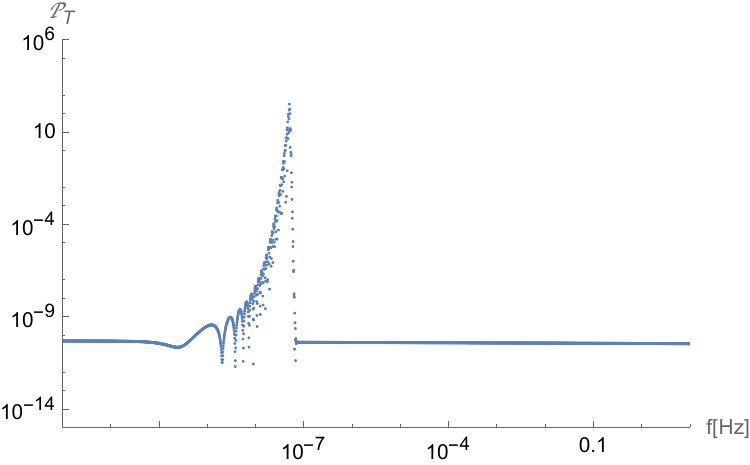}\\
  \centerline{II}
 \end{minipage}
 \begin{minipage}[t]{0.49\linewidth}
	\centering
  \includegraphics[width=.99\textwidth]{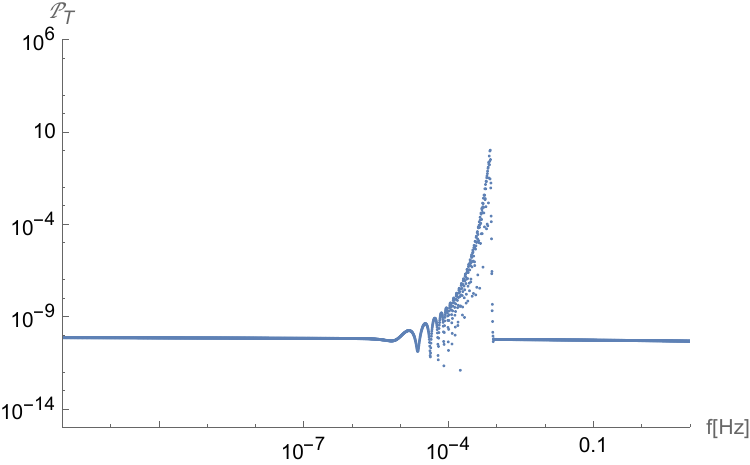}\hfill
  \centerline{III}
 \end{minipage}
 \begin{minipage}[t]{0.49\linewidth}
	\centering
  \includegraphics[width=.99\textwidth]{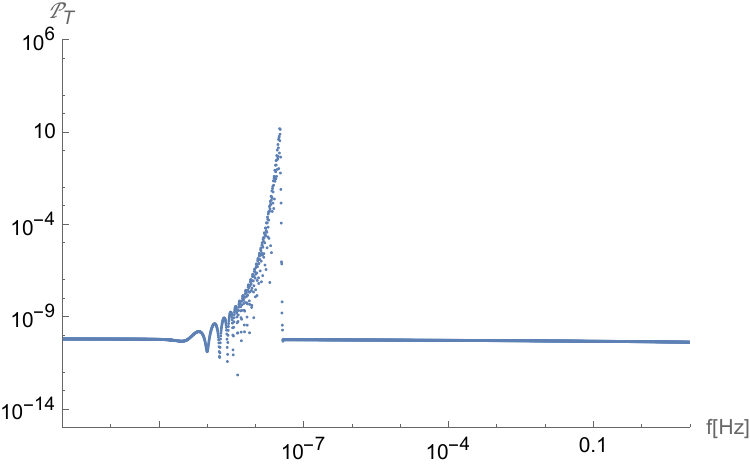}\\
  \centerline{IV}
 \end{minipage}
\caption{The tensor power spectrum for the four parameter sets of Table I.}
\label{fig:1}
\end{figure}
As shown in the figure, a peak together with oscillation will occur near the step point $\phi_c$, which originates from the rapid decline of the scalar field $\phi$ and the consequent  sudden change of the tensor sound speed $c_T^2$  during the GB term domination near the step point. The sudden change of sound speed will cause certain modes with wave numbers $k$ to cross the horizon multiple times, thus changing the phase of the mode function and generate interference, thereby resulting in oscillation in the power spectrum.
We will see in the next section that such peak correspond to an observable peak in the GW energy spectrum today.


In addition, on the CMB scale, consider the slow-roll approximation to the first order, the spectral indices $n_s$ and the tensor-to-scalar ratio $r$ can be expressed using the slow-roll parameters as
\begin{align}
n_{s} &\simeq 1-2\epsilon_1 - \frac{2\epsilon_1\epsilon_2-\delta_1\delta_2}{2\epsilon_1-\delta_1},
\end{align}
and
\begin{align}
r \equiv \frac{{\cal P}_{T}}{{\cal P}_{\mathcal{R}}} &\simeq 8|2\epsilon_1-\delta_1|.
\end{align}
The numerical results  for the four parameter sets are show in Table II.
\begin{table}
\centering
\begin{tabular}{|c||c|c|c|c|c|c|c|}
\hline
Sets&$n_s$&$r$&$\ln(10^{10}A_s)$&$ N_e$\\
\hline
I&$0.9630$&$0.019$&$3.0445$&$60.3$\\
II&$0.9675$&$0.023$&$3.0444$&$62.2$\\
III&$0.9685$&$0.036$&$3.0443$&$58.6$\\
IV&$0.9682$&$0.030$&$3.0443$&$57.7$\\
\hline

\end{tabular}
\caption{The numerical results  for the four parameter sets of Table I.}
\end{table}
We can see that the results are all consistent with the constraints from Planck 2018 $n_s = 0.9649\pm0.0042$,
$r < 0.064$ and $\ln(10^{10}A_s) = 3.044\pm 0.014 $\cite{Planck:2018jri}.

\section{Energy spectrum of GWs}

The present GW energy density spectrum $\Omega_{\rm GW,0}$ is related to the tensor power spectrum $\mathcal{P}_{T}$ obtained in the previous section as
\begin{align}\label{eqn:GWspectrum}
    \Omega_{\rm GW,0}(k)\equiv \frac{3H_0^2}{8\pi G}\frac{d\rho_{\rm GW}}{\ln k} =
    \frac{1}{12}\left(
    \frac{k}{a_0 H_0}
    \right)^2
    T^2(k)
    \mathcal{P}_{T}
    \,,
\end{align}
with the transfer function $T(k)$ represent the standard thermal history of the universe after inflation, and it can be expressed as \cite{Kawai:2023nqs,Guzzetti:2016mkm, Kuroyanagi:2014nba,Kuroyanagi:2020sfw,Boyle:2005se}
\begin{align}
    T^2(k) &=
    \Omega_m^2
    \left(
    \frac{g_*(T_k)}{g_{*,0}}
    \right)
    \left(
    \frac{g_{*s,0}}{g_{*s}(T_k)}
    \right)^{4/3}
    \left(
    \frac{3j_1(k\tau_0)}{k\tau_0}
    \right)^2 T_1^2(k) T_2^2(k)T_\nu^2,
    \label{eqn:transfer_function}
\end{align}
where $\Omega_m$ denotes the matter energy density parameter at the present time, $T_k$ denote the temperature when $k$ mode re-enters the horizon, and $j_1$ is the first spherical Bessel function. $g_*$ and $g_{*s}$ denotes the effective relativistic degrees of freedom and its counterpart for the entropy, respectively, and their changes with the temperature $T_k$ can be estimated by the following fitting function\cite{Kuroyanagi:2020sfw}
 \begin{align}
    g_*(T_k) &=
    g_{*,0}\left(
    \frac{A+\tanh\left[
    -2.5\log_{10}\left(
    \frac{f}{2.5\times 10^{-12}\, {\rm Hz}}
    \right)
    \right]}{A+1}
    \right)
        \left(
    \frac{B+\tanh\left[
    -2.0\log_{10}\left(
    \frac{f}{6.0\times 10^{-9}\, {\rm Hz}}
    \right)
    \right]}{B+1}
    \right),
\end{align}
where the frequency $f=k/(2\pi)$  and
\begin{gather}
    A = \frac{-1-10.75/g_{*,0}}{-1+10.75/g_{*,0}}
    \,,\quad
    B = \frac{-1-g_{\rm max}/10.75}{-1+g_{\rm max}/10.75}
    \,.
\end{gather}
 Similarly, $g_{*s}(T_k)$ can be obtained by replacing $g_{*,0}$ with $g_{*s,0}$ in (4.3).
 The value at the present time are $g_{*,0} = 3.36$ and $g_{*s,0} = 3.91$, and $g_{\rm max} = 106.75$ is the maximum value of $g_*$, where we take the standard model of particle physics.

The two functions $T_1^2(k)$ and $T_2^2(k)$ in (4.2) are the changes of the spectral shape due to the radiation-matter equality and reheating, respectively, which are fitting as\cite{Kuroyanagi:2020sfw}
\begin{eqnarray}
T_1^2(k)=&\left[
    1+1.57\left(\frac{k}{k_{\rm eq}}\right)+3.42\left(\frac{k}{k_{\rm eq}}\right)^2
    \right],
    \nonumber\\
 T_2^2(k)=&\left[
    1-0.22\left(
    \frac{k}{k_{\rm reh}}
    \right)^{1.5}+0.65\left(
    \frac{k}{k_{\rm reh}}
    \right)^2
    \right]^{-1}.
\end{eqnarray}
with the wavenumber corresponding to the matter-radiation equality and reheating are
\begin{align}
    k_{\rm eq} &=
    7.1\times 10^{-2}
    \Omega_m h^2
    \,{\rm Mpc}^{-1}
    \,,\\
    k_{\rm reh} &=
    1.7\times 10^{14}
    \left(
    \frac{g_{*s}(T_{\rm reh})}{106.75}
    \right)^{1/6}
    \left(
    \frac{T_{\rm reh}}{10^7\,{\rm GeV}}
    \right)\,{\rm Mpc}^{-1}
    \,,
\end{align}
where we take the reheating temperature $T_{\rm reh}=10^4$ GeV for the standard reheating scenario.

The last term $T_\nu$ in the transfer function (4.2) is the damping effect arising from the free-streaming of neutrinos, which can be fitting below the neutrino decoupling frequency as \cite{Boyle:2005se}:
\begin{align}
    T_\nu &=
    \frac{15}{343(15+4f_v)(50+4f_v)(105+4f_v)(108+4f_v)}
    \nonumber\\
    &\quad\times
    \big(
    14406 f_v^4 - 55770 f_v^3 + 3152975 f_v^2 - 48118000 f_v + 324135000
    \big)
    \,,
\end{align}
with $f_v=0.4052$ is the energy density fraction of the neutrinos.

In Fig.4, we show the spectrum of GWs at the present time as a function of frequency. The curves in the upper part represents the sensitivity curves of some current or planned detectors~\cite{GWob1,GWob2,GWob3,GWob4,GWob5,GWob6}, and the constraints of EPTA(orange region)\cite{EPTA1,EPTA2} and NANOGrav(green region)\cite{NANOGrav:2023gor,NANOGrav:2023hde} are also shown there.

\begin{figure}[htb]

 \begin{minipage}[t]{0.49\linewidth}
\centering
  \includegraphics[width=.99\textwidth]{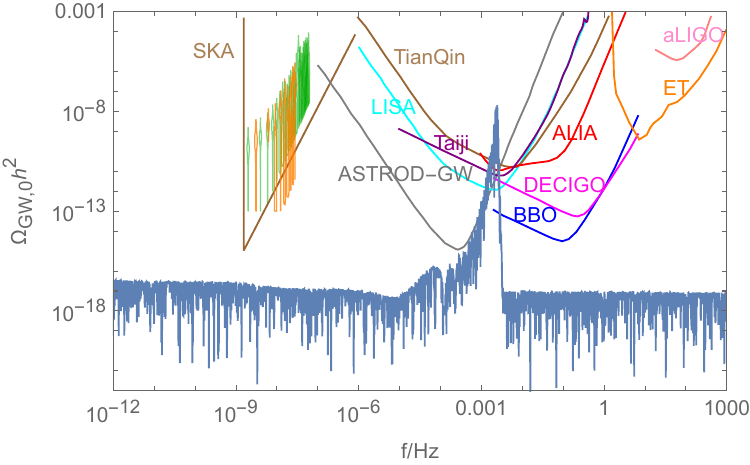}\hfill
  \centerline{I}
 \end{minipage}
 \begin{minipage}[t]{0.49\linewidth}
	\centering
  \includegraphics[width=.99\textwidth]{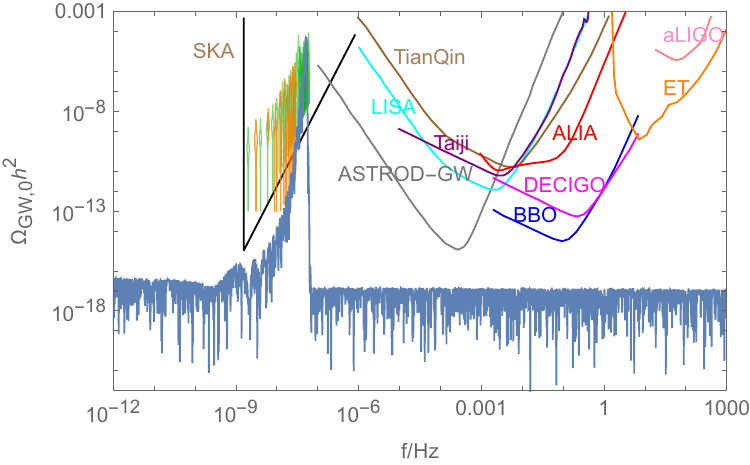}\\
  \centerline{II}
 \end{minipage}
 \begin{minipage}[t]{0.49\linewidth}
	\centering
  \includegraphics[width=.99\textwidth]{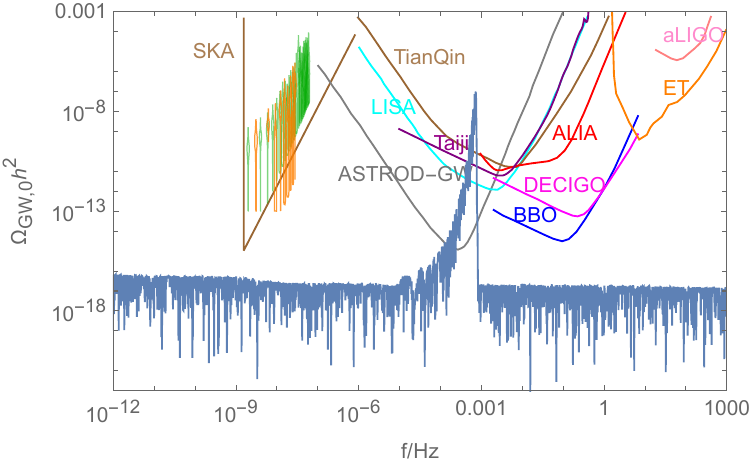}\hfill
  \centerline{III}
 \end{minipage}
 \begin{minipage}[t]{0.49\linewidth}
	\centering
  \includegraphics[width=.99\textwidth]{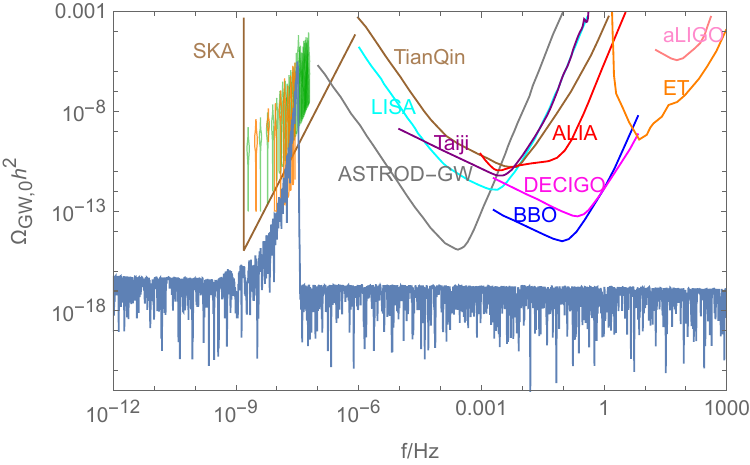}\\
  \centerline{IV}
 \end{minipage}
\caption{The  GWs energy spectrum for the four parameter sets of Table I.}
\label{fig:1}
\end{figure}

We can see that the GW energy spectrum also exhibits a peak corresponding to the peak in the tensor power spectrum. For parameter sets I and III, the peak frequency is approximately $10^{-3}$Hz, with the curves lies above the expected sensitivity curves of LISA, Taiji and TianQin, this implies detectability by future space-based GW detectors. While the parameter sets II and IV yield a peak within the nanohertz frequency range, which can be used to explain the latest PTA observations.

\section{The case with double peaks}

In this section, we analyze the coupling coefficient $\xi$ modeled by two hyperbolic tangent functions as
\begin{eqnarray}
\xi=\xi_{01}\tanh[\xi_1(\phi-\phi_{c1})]+\xi_{02}\tanh[\xi_2(\phi-\phi_{c2})].
\end{eqnarray}
The motivation is that if we assume that the moduli space has multiple minima, allowing for two or more wall-crossing processes of $\phi$. So the coupling coefficient $\xi(\phi)$ can take the form of a multi-step function. Here we assume that there are two step-like functions, then the scalar field will experience two rapid decline stages near the step points $\phi_{c1}$ and $\phi_{c2}$, correspondingly. These dynamics generate a double-peak structure in the tensor perturbation power spectrum,  which will lead to double-peaks within the GW energy spectrum.

We take the parameters in Table III as an example.
\begin{table}
\centering
\begin{tabular}{|c||c|c|c|c|c|c|c|c|c|}
\hline
Sets&$n$&$V_0$&$\sqrt{\frac{2}{3\alpha}}$&$\xi_{01}$&$\xi_{02}$&$\xi_1$&$\xi_2$&$\phi_{c1}$&$\phi_{c2}$\\
\hline
V&$1/4$&$1.174 \times10^{-9}$&$0.02$&$9.2\times10^{8}$&$10.1\times10^{8}$&$2.4$&$2.4$&$8.86$&$6.50$\\
\hline

\end{tabular}
\caption{Example of the parameter sets.}
\end{table}
And the evolution  of the scalar field $\phi$ and $c^2_T$ as a function of the e-folding number $N_e$ are shown in Fig.5,
\begin{figure}
  \begin{minipage}[t]{0.49\linewidth}
	\centering
	\includegraphics[width=.99\textwidth]{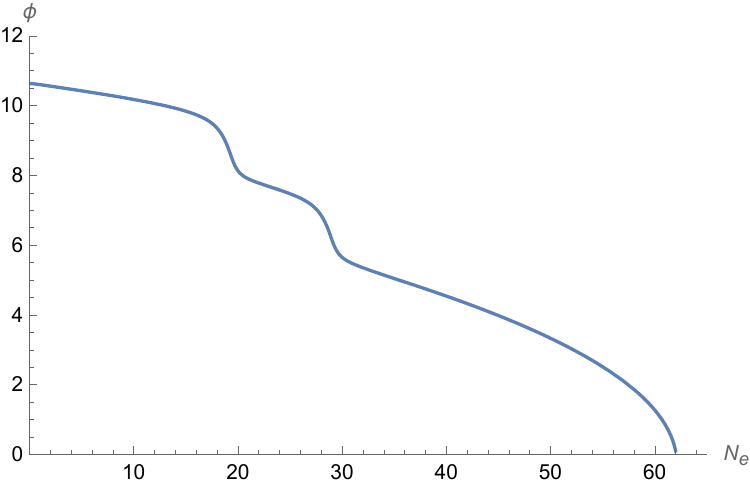}
	\label{fig:a} %
 \end{minipage}
 \begin{minipage}[t]{0.49\linewidth}
	\centering
	\includegraphics[width=.99\textwidth]{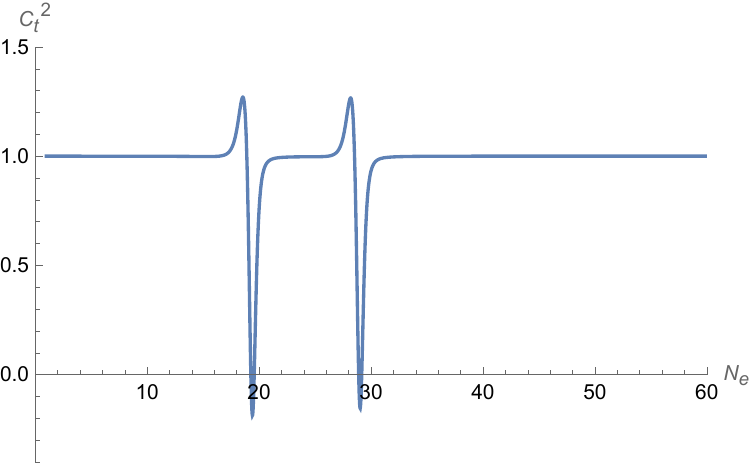}
	\label{fig:b}
 \end{minipage}
 \caption{The relation of inflaton $\phi$ and $c^2_T$ evolution with the e-folding number for the double-peak model.}
\label{fig:1}
\end{figure}

The numerical results of the tensor perturbation power spectrum are show  in Fig.6.
\begin{figure}
\centering
 \begin{minipage}[t]{0.7\linewidth}
	\centering
	\includegraphics[width=.99\textwidth]{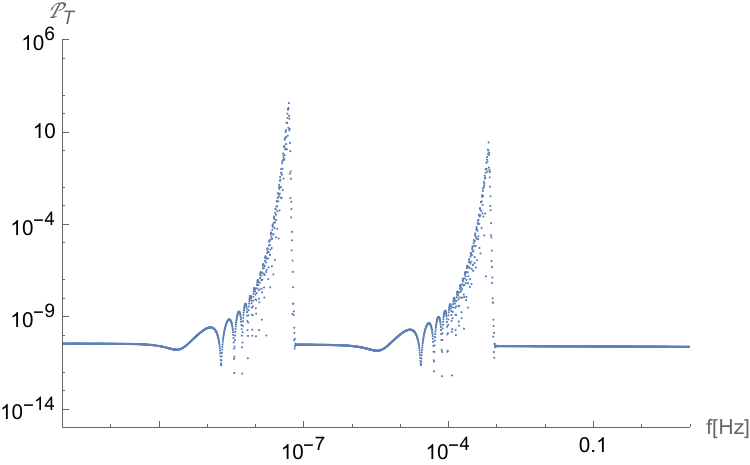}
	\label{fig:a}
 \end{minipage}
 \caption{The numerical results of the tensor power spectrum for the parameter set V.}
\label{fig:1}
\end{figure}
We can see that there are two rapid decline stages near the step point $\phi_{c1}$ and $\phi_{c2}$. Such rapid decline of $\phi$ during the GB term dominated period near the step points
will causes double peaks in the tensor power spectrum. In addition, the numerical results of dynamics on CMB scale are presented in Table IV, which are all consistent with the constraints from Planck 2018.
\begin{table}
\centering
\begin{tabular}{|c||c|c|c|c|c|c|c|}
\hline
Sets&$n_s$&$r$&$\ln(10^{10}A_s)$&$ N_e$\\
\hline
V&$0.9629$&$0.017$&$3.0448$&$62.0$\\
\hline

\end{tabular}
\caption{The numerical results  for the double-peak model.}
\end{table}

Finally, the numerical results of the GW energy spectrum is shown in Fig.7.
\begin{figure}
\centering
 \begin{minipage}[t]{0.7\linewidth}
	\centering
	\includegraphics[width=0.99\textwidth]{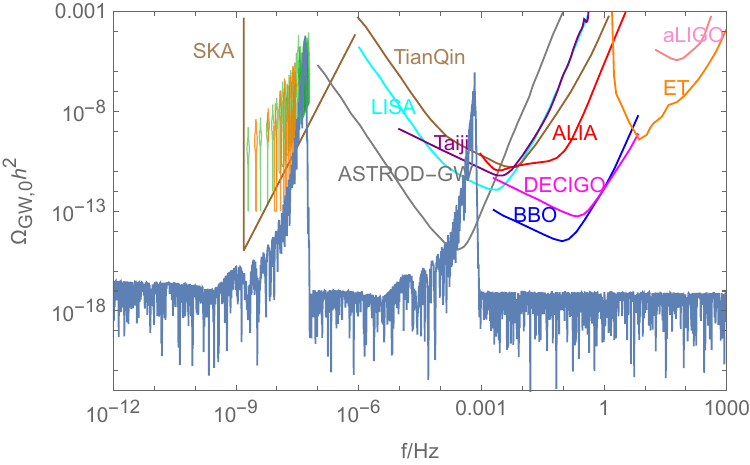}
	\label{fig:a}
 \end{minipage}
 \caption{The numerical results of the GWs energy spectrum for the parameter set V.}
\label{fig:1}
\end{figure}
As shown in the figure, the two peaks in the tensor power spectrum produce a double-peak structure in the GW energy spectrum. One peak corresponds to a frequency of $4.8\times10^{-8}$Hz, falling within the observation range of the PTA, which can explain the observation. The other peak occurs at a frequency of $6.7\times10^{-4}$Hz, with its energy spectrum exceeding the sensitivity curves of space-based GW detectors such as LISA, ALIA, Taiji, and TianQin, making it a viable target for near-future observations.

Furthermore, the dual-peak nature of the spectrum enables cross-validation across detectors. If a PTA detector identifies the nanohertz-frequency peak, the complementary millihertz-frequency peak should be observable by space-based interferometers like LISA. This contrasts with single-peak models, which can only be detected in one frequency range.  Therefore, the double-peaks model can be distinguished from other single peak models

\section{Summary}

In this paper, we discuss the generation of GWs in a model where the inflaton is non-minimally coupled to the GB term. Assuming the inflation potential $V(\phi)$ corresponds to the E-model attractor potential and the coupling coefficient $\xi(\phi)$ takes the form of a hyperbolic tangent function, which is similar to a step function. If the parameters satisfy $V_{,\phi}\xi_{,\phi}>0$, when the inflaton rolls approaches the step point, the GB term briefly dominate the inflationary dynamics.  This dominance induces an increase in the velocity of the inflaton, therefore, leading to an instantaneous rapid decline. For certain parameter spaces, $c^2_T$ temporarily becomes negative, consequently, the mode equation of tensor perturbations develops a transient growing solution, resulting in a distinct peak within the primordial tensor perturbation power spectrum.

We solve the perturbation equations numerically to obtain the tensor power spectrum, apply the transfer function representing the
standard thermal history of the Universe after inflation, and finally derive the present-day GW energy spectrum. We found that for certain parameter  spaces, e.g., sets I and III, the energy spectrum exhibits a peak around millihertz frequencies, lying above the projected  sensitivity curves of LISA, Taiji,etc, making it detectable by upcoming space-based interferometers.  For parameter sets II and IV, the peak lies within the $10^{-7}-10^{-8}$Hz range, surpassing the expected sensitivity thresholds of SKA and PTA, which can be used to explain the PTA observational signals.

Furthermore, we discuss the feasibility of generating a double-peak GW spectrum using a coupling coefficient $\xi$ modeled as  a double-step function.
In this scenario, the corresponding tensor power spectrum exhibits two distinct peaks. Through numerical calculations of the present-day GW energy spectrum, we demonstrate that with appropriate parameter sets, for example, the parameter set V, the spectrum can simultaneously feature peaks in both the nanohertz and millihertz frequency bands. These peaks can be detected by PTA in the nanohertz range and by space-based GW detectors in the millihertz range, simultaneously. Consequently,  such double-peak models can be observationally distinguished from single-peak models.

\acknowledgments

This work was supported by  "the Natural Science Basic Research Program of Shaanxi Province" No. 2023-JC-YB-072. And supported by "the Fundamental Research Funds for the Central Universities" No. ZYTS25130.



\end{document}